\documentclass{jpsj-suppl}
\usepackage{txfonts} 

\title{Neutrino Telescope Array (NTA) \\
 \vskip0.2cm\hskip1.77cm \Large --- \ Towards Survey of Astronomical $\nu_\tau$ Sources}

\author{
George W.-S.~\textsc{Hou}, 
(for the NTA Collaboration)
}

\inst{
 {Department of Physics, National Taiwan University, Taipei 10617, Taiwan}
}

\email{wshou@phys.ntu.edu.tw}

\recdate{July 27, 2014}

\abst{
The Earth-skimming $\nu_{\tau}$ method allows for
huge target mass and detection volume simultaneously.
In part motivated by IceCube PeV astro-neutrino events,
the planned NTA observatory has three site stations
watching the air mass surrounded by Mauna Loa, Mauna Kea,
and Hualalai on Hawaii Island, plus a
site station at the center watching the lower night sky.
Sensitivities equivalent to $> 100$ km$^3$ water and
pointing accuracy of $< 0.2^\circ$ can be achieved
with Cherenkov-fluorescence stereoscopic observation for
PeV-EeV neutrinos that is almost background-free.
With design based on experience from Ashra-1 and
the goal of clear discovery and identification of
astronomical $\nu_\tau$ sources,
a new international collaboration is being formed.}

\kword{Earth-skimming $\nu_{\tau}$, PeV-EeV $\nu$s, Ashra-1, NTA, pointing accuracy,
      collaboration}

\begin{document}
\maketitle

\section{Introduction: Earth-skimming $\nu_\tau$ Method}

In Sugiyama san's opening introduction, there were a few slides on
``New Concepts from VHEPA-3'' from the predecessor meeting held more than a decade ago,
which included the NuTel Project~\cite{NuTel}, and on
``\textbf{A}ll-sky \textbf{S}urvey \textbf{H}igh \textbf{R}esolution
\textbf{A}ir-shower detector (Ashra)''~\cite{Ashra}.
Here we report on the Neutrino Telescope Array, or NTA,
which can be said as the joining of forces and concept of Ashra and NuTel,
and plans to use Ashra-1 technology for
\begin{verse}
\underline{\ Aim/Scientific Goal\ }:
\ Clear Discovery and Identification of \\
\hskip3.6cm Nonthermal Hadronic Processes in the Universe, \\
\hskip3.6cm be it Galactic, Extragalatic, or Cosmogenic.
\end{verse}

It is quite amazing to me that, when pushing the NuTel project
more than a dozen years ago, we highlighted the ``window of opportunity''
between PeV and EeV as lacking developed detector techniques.
With the successful completion of IceCube and Auger,
with capabilities up to PeV range and above EeV range, respectively,
this window remains largely open.
Furthermore, IceCube entices us with a glimpse~\cite{IceCube2013} of
PeV astro-neutrino events!

What caught my attention back then was the Earth-skimming and
mountain-pentrating $\nu_\tau$ detection method.
As cross sections grow with energy,
very high energy (VHE) $\nu_\tau$s can convert in the Earth or a mountain,
the emerging $\tau$s then shower in the atmosphere,
hence allow for both very large effective target mass and detection volume.
In contrast, for $\nu_e$, the electron shower is absorbed in mountain/Earth,
while the converted muon e from $\nu_\mu$ does not shower.
This Earth-skimming $\nu_\tau$ technique, with sensitivity in the PeV-EeV range,
can probe hadron acceleration in astronomical objects,
whereby $\nu_\mu$ oscillates to $\nu_\tau$ along the way.
An actual UV telescope placed at a distant (several 10 km) mountain
can catch the Cherenkov pulse from the $\tau$ shower (hence the name ``NuTel'').
There are further advantages of the method: mountain/Earth as shield of cosmic ray secondaries,
precise arrival direction determination, and negligible atmospheric neutrino background.

In the following, we recap briefly the NuTel effort and its demise,
describe (material courtesy Makoto Sasaki) the Ashra-1 detector
that made the first search for GRB $\nu_\tau$s
utilizing the Earth-skimming method,
then turn to NTA, a new collaboration to-be that
aims seriously at discovering $\nu_\tau$ sources
in the PeV-EeV range.

\section{``My'' NuTel Effort}

In 2001, when I was on the look-out for a particle astrophysics project,
Francois Vannucci came by for a visit. I was quite intrigued by what
he called the ``Mountain-Valley $\nu_\tau$ Detection''.
I asked him whether he already had funding ...
After checking literature, and discussing with faculty members in our group,
we added some manpower along the direction.

Interestingly, the Hawaii site also came out from Vannucci's visit.
I remember bringing Vannucci to a courtesy visit of
Fred Lo, the Director of ASIAA at the time, and PI of the ``CosPA-1''
project that was building the AMiBA microwave array
(I lead the PA part, called CosPA-2).
Well, AMiBA was to be built on Mauna Loa.
Chatting with Fred Lo in front of a map of Haiwaii Island,
both Vannucci and myself went \emph{gotcha} at the same time:
(together with Mauna Kea) two big mountains with 40 km separation!
Subsequently, it was Alfred Huang who pointed out
Mt. Hualalai as a site with view of the broad side (90 km wide) of Mauna Loa.

The NuTel project \cite{Velikzhanin:2005rs} was launched,
with the aim of observing Cherenkov radiation from
$\nu_\tau$-originated air showers.
This was meant as a ``speedboat'' approach,
to place a telescope up Mauna Loa (or Hualalai) to
watch Mauna Kea (or Mauna Loa), hence was more ``mountain penetrating''.
There are three stages for the simulation:

\ \ \ (1) $\nu_\tau \to \tau$ conversion in mountain;

\ \ \ (2) $\tau$ shower development;

\ \ \ (3) detector performance.

\noindent For reconstruction, it was deemed necessary to
place two telescopes $\sim 100$ m apart for stereo view,
and utilizing a cluster-based trigger algorithm.
It was found that angular error would be $< 1^\circ$,
shower energy resolution was decent,
and reconstruction efficiency was good when triggered.

The MAPMT-based readout electronics up to a DAQ system was developed,
even built, in a year's time. The bottleneck was optics.
We checked around and found EUSO-like Fresnel lens system not practical,
but I myself already felt confident enough to go up Mauna Loa
for a site visit in beginning of 2004, having filed for
the 4-year renewal of CosPA-2 in Fall 2003.

NuTel is the first experiment dedicated to Earth-skimming $\tau$ appearance,
but it did not reach physics running.
Its Achilles heel was the estimated event rate of 0.5 event/year.
Because of this, the proposal was rejected in Spring 2004,
and we could not restore it, after several tries.
But I continued it on a shoestring ...
We were brave enough to develop our own optics,
eventually building not one, but two 1.8 m Schmidt mirror systems.
We brought the first one up Mei-Fong at 2100 m in Taiwan, Summer 2009,
to watch 3000 m peaks at night. Through this we learned
the difficulty of mountain operation.
Together with funding issues, we sought synergies with
CRTNT effort in Beijing, but the group had evolved.
So, the NuTel project was effectively terminated around 2010.

Until we got contacted by Ashra-1 two years laters.

\section{Ashra-1:  $1^{\rm st}$ Search for GRB $\nu_\tau$
 \hskip5cm {\normalsize\rm (courtesy Makoto Sasaki)}}

Ashra-1 and NuTel were in contact back in 2003-2004.

The All-sky Survey High Resolution Air-shower detector~\cite{Ashra} Phase I, or Ashra-1,
optimized to detect VHE particles, was developed slightly earlier but
in a similar time frame as NuTel, with the more ambitious aim of
``multi-messenger astronomy''~\cite{Barwick00,Sasaki00}.
Starting from ultra wide optical system with 42$^{\circ}$ field-of-view (FOV),
it demagnifies to 1 inch at focal surface by using photon and electron optics~\cite{PLI11}.
Combined with a high resolution imaging system with trigger,
very cost-effective pixels compared with
conventional photomultiplier arrays are achieved.
Ashra-1 can observe the entire sky with arcminute resolution,
with each of its 12 Detector Units consisting of several aligned Light Collectors (LC).
What is important for our current topic is that
Ashra-1 succeeded in demonstrating the power of the Earth-Skimming (ES) $\nu_\tau$ Method,
and our aim now is to realize the NTA.

Let us discuss a little more about the Ashra-1 detector design.
The Ashra-1 LC has 42$^\circ$ field of view,
with total resolution of $\sim 3$ arcmin.,
and can cover Mauna Kea surface at 35 km distance (i.e. from Mauna Loa).
Ashra-1 takes a multi-messenger approach with one detector system.
The key technical feature of Ashra-1 is the use of electrostatic
rather than optical lenses to generate convergent beams
(the 20 inch Photoelectric Lens Image (PLI) tube~\cite{PLI11}),
enabling high resolution over a wide FOV.
The electron optics of photoelectric lens imaging
links wide angle precision optics~\cite{2002NIMPA.492...49S}
to the image pipeline~\cite{2003NIMPA.501..359S}.
The Photoelectric Image Pipeline (PIP) splits the focal image
into trigger/image capture devices after amplification,
sending the same fine image to multiple triggers.
This allows the simultaneous measurement of three phenomena
on different time scales, i.e. Cherenkov emission (ns), fluorescence ($\mu$s),
and starlight (s), without sacrificing the S/N ratio.

The demonstration phase has been running since 2008 at the
Mauna Loa observation site (ML-OS) at 3300 m above sea level on Hawaii Island.
With GRB alert given by SWIFT, Ashra-1 succeeded in the
first search for PeV-EeV $\nu_\tau$s originating from a GRB~\cite{AshraCNeu}
with the Earth-skimming $\nu_{\tau}$ technique.
Moreover, Ashra-1 has achieved the best-yet instantaneous sensitivity
in the 100~PeV energy region subsequent to a January 2012 trigger upgrade.

But Ashra-1 also ran out of ways and means,
so let us now unfold the NTA project,
based on the achievements of Ashra-1 and NuTel.

\section{NTA: a New Collaboration}

Based on Ashra-1 performance, we aim at forming a new collaboration,
Neutrino Telescope Array (NTA),
with the stated scientific goal:
\emph{clear discovery and identification of
nonthermal hadronic processes in the Universe,
be it Galactic, extragalatic, or cosmogenic.}

Stimulation has come from the IceCube experiment,
which has recently reported~\cite{IceCube2013}
the ``first indication of an astrophysical neutrino flux''.
They observe two (with one more subsequently added) fully contained
$\nu$-induced particle showers with deposited energy $\sim$ 1 PeV,
which provides great motivation for exploring the PeV-EeV region.
\begin{center}
  --- \emph{What if one had better Sensitivity
 and accurate Pointing?} ---
\end{center}
With better than $0.2^\circ$ pointing accuracy,
NTA would be able to discern the origins of such events.

\begin{figure}[t]
\begin{center}
\begin{tabular}{cccc}
\begin{minipage}[t]{0.39\linewidth}
\includegraphics[width=\linewidth]{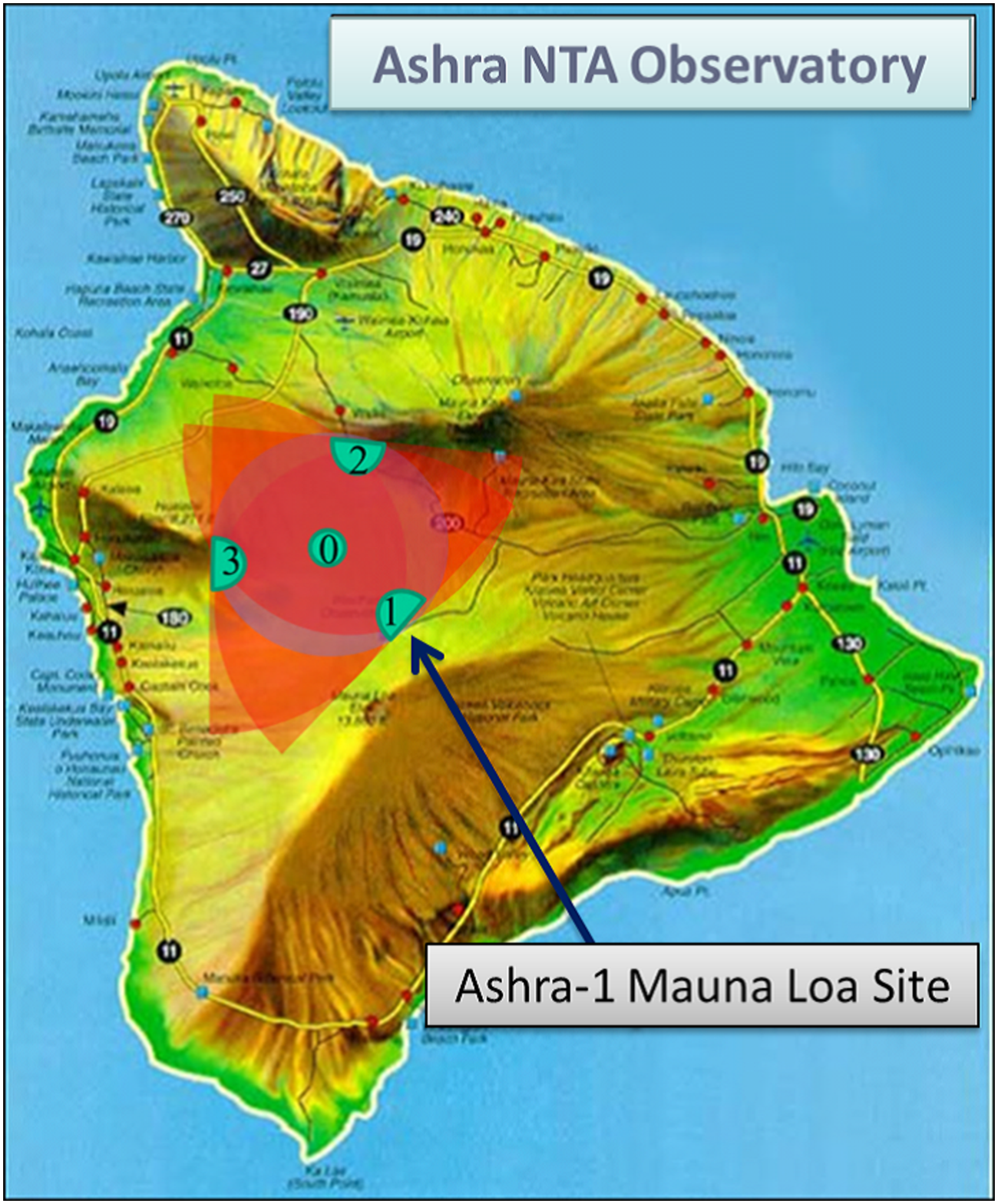}
\end{minipage} &
\hskip0.3cm
\begin{minipage}[t]{0.47\linewidth}
\includegraphics[width=\linewidth]{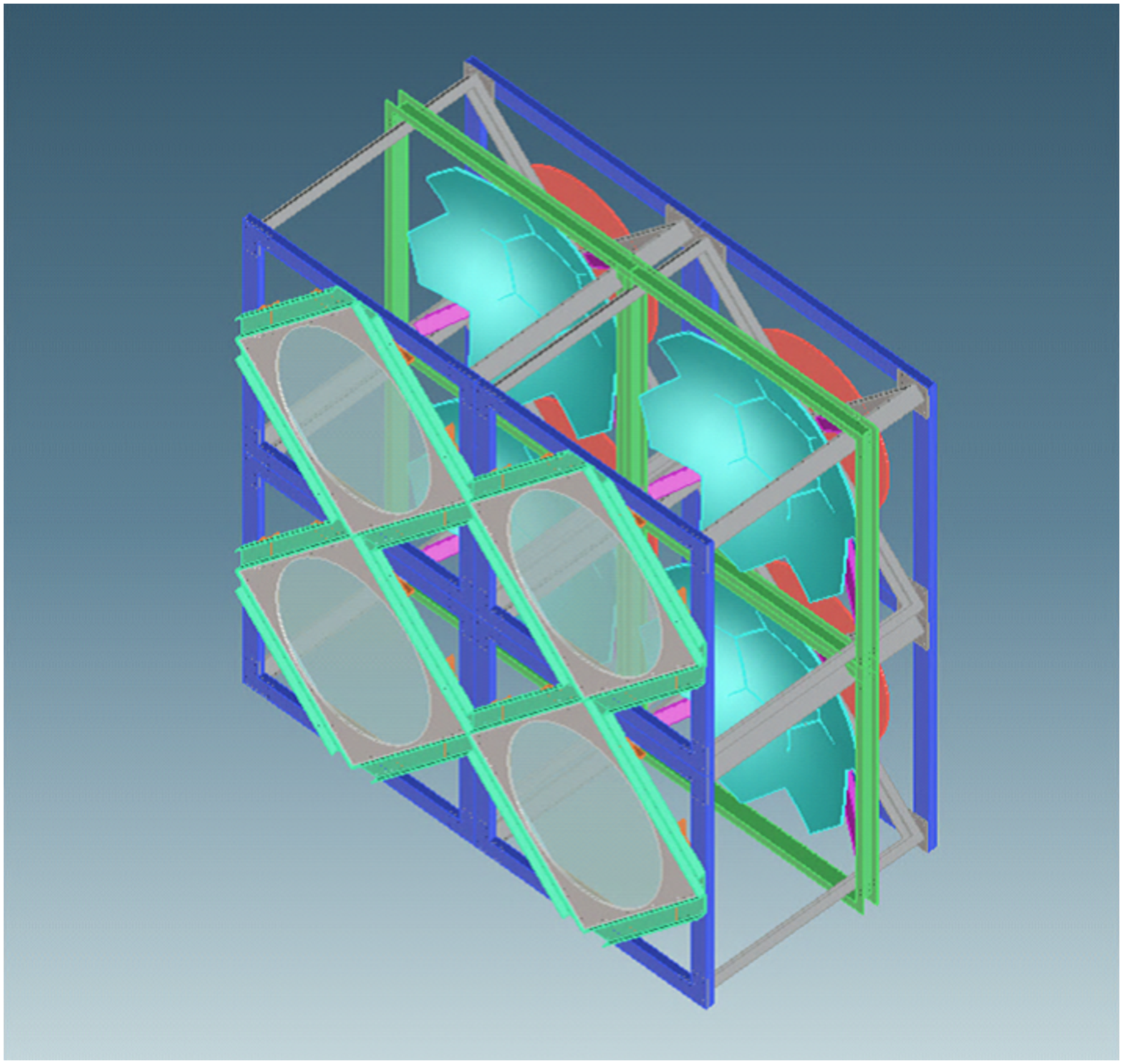}
\end{minipage}
\end{tabular}
\caption{
 (left) NTA observatory layout on Hawaii Island, where overlay of semi-circles
  illustrate view from Site1--3;
 (right) Detector Unit of four same Light Collectors.
}
\label{fig:NTA}
\end{center}
\end{figure}

\begin{table}[b]
\caption{
The x-y-z coordinates and FOV coverage of the NTA sites used in the simulation,
where the z-axis points to zenith and y-axis points north,
and z-coordinates are from topography data.
Site1--3 form an equilateral triangle, with Site0 at the geometric center
and defined as the origin.
Site1 is located at ML-OS (on Mauna Loa) and Site2 at 25~km distance from ML-OS
in the direction of the Kilohana Girl Scout Camp.
}
\label{tab:site}
\begin{center}
\begin{tabular}{clrrcc}
\hline
Site ID & Location & X [km] & Y [km] & Z [km] & FOV [sr] \\
\hline
Site0 & Center    &   0.000 &   0.00 \ \ &    2.03 &  $\pi$	\\
Site1 & Mauna Loa &   9.91 \ \ & $-10.47$ \ \ &    3.29 &  $\pi/2$ \\
Site2 & Mauna Kea &   4.12 \ \ &  13.82 \ \ &    1.70 &  $\pi/2$ \\
Site3 & Hualalai  & $-14.02$ \ \ &  $-3.35$ \ \ &    1.54 &  $\pi/2$ \\
\hline
\end{tabular}
\end{center}
\end{table}

\vskip0.2cm
\noindent\underline{Plan and Size}
\vskip0.1cm

As shown in Fig.~\ref{fig:NTA} (left), the planned NTA observatory
consists of four sites: Site0--3, where the x-y-z coordinates
and corresponding FOV coverage are given in Table~\ref{tab:site}.
Site1--3 are the vertices of an equilateral triangle of 25~km sides,
and observe the total air mass surrounded by Mauna Kea, Mauna Loa, and Hualalai;
the central Site0 can potentially have full-sky coverage.



VHE $\nu_\tau$s can convert in Earth/mountain and reappear as
$\tau$s and produce air showers upon decay in the atmosphere,
and the resulting Cherenkov photons are detected~\cite{Fargion02, Feng02, Hou02}.
Owing to separation of $\nu_{\tau} \to \tau$ conversion
from subsequent air shower generation, detection is possible
while preserving the huge target mass required for the initial interaction.
%
Currently, Ashra-1 operates on Site1, or ML-OS, with view of
Mauna Kea, which is a huge mountain equivalent to $10^4$ km$^3$ of ice
for converting $\nu_\tau$ to $\tau$, but it also serves as a shield to CR background.
The distance between Mauna Kea and Loa allows a 30 km range for shower development.
It is with this configuration that Ashra-1 has demonstrated~\cite{AshraCNeu}
the ES-$\nu_\tau$ technique, but NTA would augment it
with the fluorescence ability of Site0, plus two other mountain sites.
The huge target mass ($> 100$ km$^3$),
huge atmospheric mass (shower volume, with area $> 1000$ km$^2$)
and mountain as background shield
imply a rather large footprint for NTA.

Each site has a group of Detector Units (DU), each of which has
4 LC systems (Fig.~\ref{fig:NTA} (right)) instrumented with segmented mirrors.
The NTA LC concept is scaled up from Ashra-1 by 1.5, but uses the same trigger and readout.
The LCs use Schmidt optics, with pupil of 1.5m,
4 LCs with same FOV give a DU with effective pupil size of 3m.
12 DUs are needed per $\pi$ solid angle coverage.
Thus, from Table~\ref{tab:site}, at least 30 DUs are needed.
The construction, deployment and operation of these DUs
spread out at four distant sites defines the need for
NTA to be an International Collaboration.

\vskip0.2cm
\noindent\underline{Pointing Accuracy}
\vskip0.1cm

Pointing accuracy is one of the main strengths of NTA.
We have studied NTA performance based on Ashra-1 experience,
and a Letter of Intent (LOI) is at hand~\cite{LOI}.
Detailed design studies for the NTA detector are currently underway.
%
%
%

We estimate our ability to trace showers back to their origins,
which is a very important feature in light of the IceCube events.
The steps of simulation are very similar to the one mentioned
for NuTel, except that one also has the fluorescence ability of Site0.
Light propagation and quantum efficiency are better, and
the highlight is the much higher pixelization of Ashra-1 detector development. 

First, we use PYTHIA to model neutrino interactions.
The $\tau$ with respect to the parent $\nu_{\tau}$ angle,
$\Delta \theta_{\tau}$, is less than 0.3~arcmin. for $E_{\tau}>1$~PeV.
The Ashra NTA detector design is optimized for this.

Second, we use GEANT4 to evaluate the deflection of $\tau$ as it
propagates within the Earth, adopting the parametrization of \cite{Dutta2001} to estimate energy loss, where radiative energy loss is dominant.
Bremsstrahlung, $e^+e^-$ pair production, and photonuclear interactions are all included.

Next, to estimate the deflection due to $\tau$ decay,
we use TAUOLA and take account of $\tau$ polarization.
The deflection angle is less than 1 arcmin.
if the energy of the secondary particle is higher than 13~TeV.
TAUOLA showed that the probability of a deflection greater than 1 arcmin.
for $\tau$s with PeV energy is very small.
We adopt this assumption throughout our air-shower analysis.

Finally, the hadron air-shower direction is evaluated using CORSIKA.
We compare the direction of the parent particle (charged pion) at shower max
to that of $e^\pm$, the dominant producers of Cherenkov photons.
The angle between the average direction of $e^\pm$ and the parent particle
is found to be within $0.1^\circ$ at 1~PeV.
The energy resolution is found, by MC simulation, to be several 10\%.

We conclude that the arrival direction of PeV-scale $\nu_{\tau}$s is
within $0.1^\circ$ of the original direction of the generated hadron air-shower.
The accurate reconstruction of arrival direction by means of fine imaging will be
a very powerful technique in the determination of point sources of PeV $\nu_{\tau}$s.

\vskip0.2cm
\noindent\underline{Performance and Sensitivity}
\vskip0.1cm

To simulate the performance of the Ashra NTA detector,
we assume each DU has $32^{\circ} \times 32^{\circ}$ total FOV,
$0.5^{\circ} \times 0.5 ^{\circ}$ for trigger pixel FOV,
and $0.125^{\circ} \times 0.125^{\circ}$ image sensor pixel FOV.
According to Table~\ref{tab:site}, the Site0 system consist of
12 DUs, which covers the solid angle of $\pi$~sr in the
lower elevation angle regions (can be extended to full-sky coverage).
The remaining sites have only 6 DUs in the
lower elevation angle region covering $\pi/2$~sr.
The bottom edge of the lower elevation angle region is
defined to be $-9^{\circ}$ (below the horizon).


%
%

In our simulation program, we take density profile of the Earth,
 use the $\nu_{\tau}$ distribution from CTEQ4~\cite{Gandhi98},
 inelasticity parameter from~\cite{Gandhi96},
 and parameterize energy loss in Earth by ~\cite{Dutta2001, Tseng03}.
We use $\tau$ decay from TAUOLA and
 air-shower generation of Gaisser-Hillas $+$ NKG~\cite{TANeu}.
We use a constant average $\nu_{\tau}$ energy fraction of 40\% (lab frame) from $\tau$ decays,
without taking into account the energy distribution.
The error from this approximation is found to be negligible.
For detector simulation, we incorporate light collection and throughput
with simplified triggering logic. Event reconstruction is not yet implemented.
All candidate events must satisfy the trigger conditions:

\ \ \ (1) number of detected photoelectrons per LC $>$ 61;

\ \ \ (2) $S/N$ estimated in track-associated $4\times 64$ pixel box
(air-shower track included) $>$ 4~\cite{Sasaki2001}.

\begin{figure}[b]
\begin{center}
\begin{tabular}{cc}
 \begin{minipage}[t]{0.37\hsize}
  \begin{center}
    \includegraphics[width=\hsize]{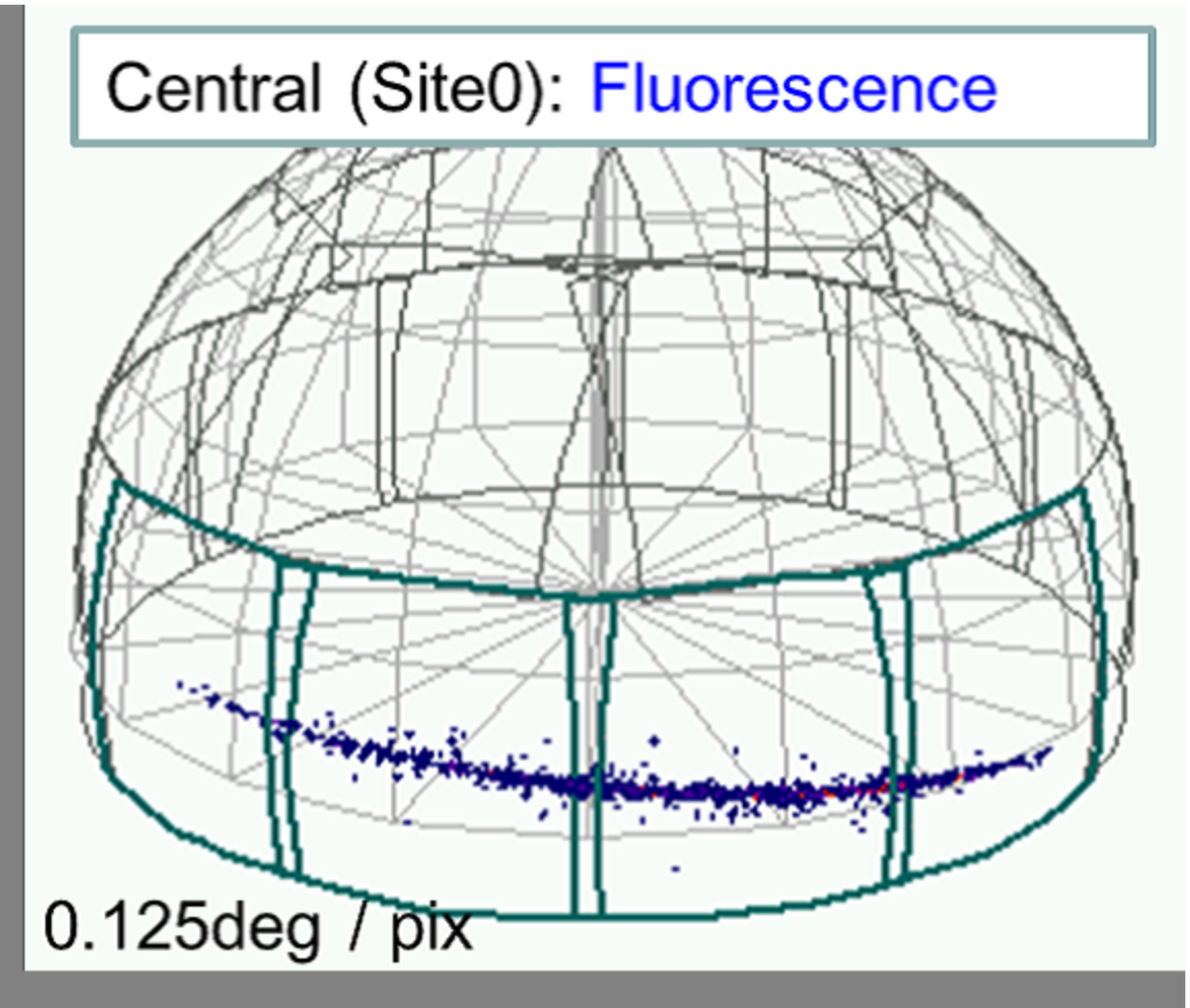}
  \end{center}
 \end{minipage} &
  \hspace{7mm}
 \begin{minipage}[t]{0.37\hsize}
  \begin{center}
    \includegraphics[width=\hsize]{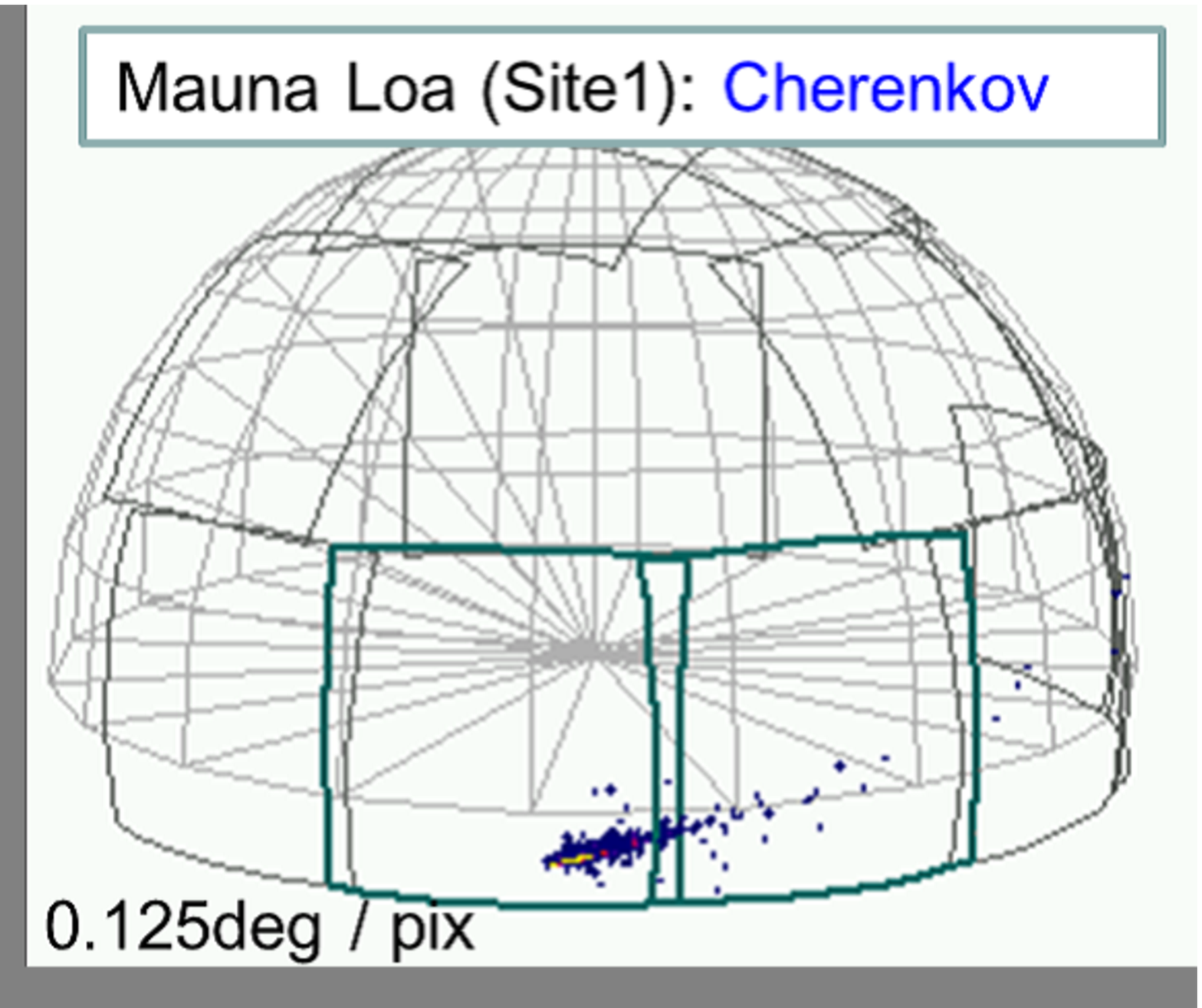}
  \end{center}
 \end{minipage} \\
\end{tabular}
\caption{
Imaging by Site0 (left) and Site1 (right)
for simulated event with $E_{\nu_\tau}=10^{17}$ eV.
}
\label{fig:SimEvt}
\end{center}
\end{figure}

\noindent A simulated event with primary $\nu_{\tau}$ energy $E_{\nu_\tau}=10^{17}$eV,
elevation angle $-6.4^\circ$ and arrival direction towards Mauna Loa,
consistent with the above conditions, is shown in Fig.~\ref{fig:SimEvt}.
A combined simple fit to Site0$+$Site1 gives an
error for $\nu_{\tau}$ arrival direction reconstruction at 0.08$^{\circ}$.

\begin{figure}[t!]
\begin{center}
\begin{tabular}{cccc}
\begin{minipage}[t]{0.45\linewidth}
\includegraphics[width=\linewidth]{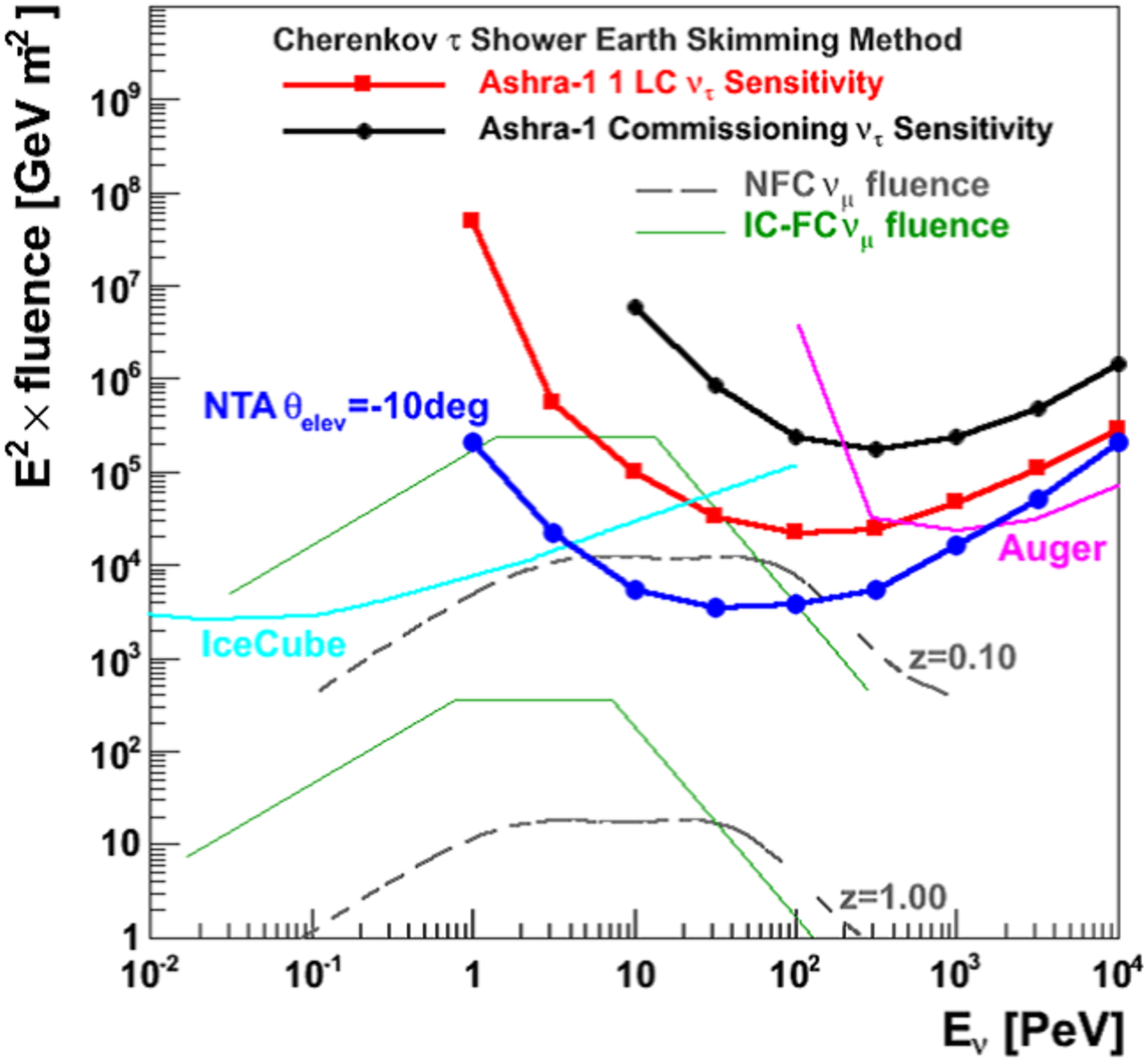}
\end{minipage} &
\hskip-0.3cm
\begin{minipage}[t]{0.56\linewidth}
\includegraphics[width=\linewidth]{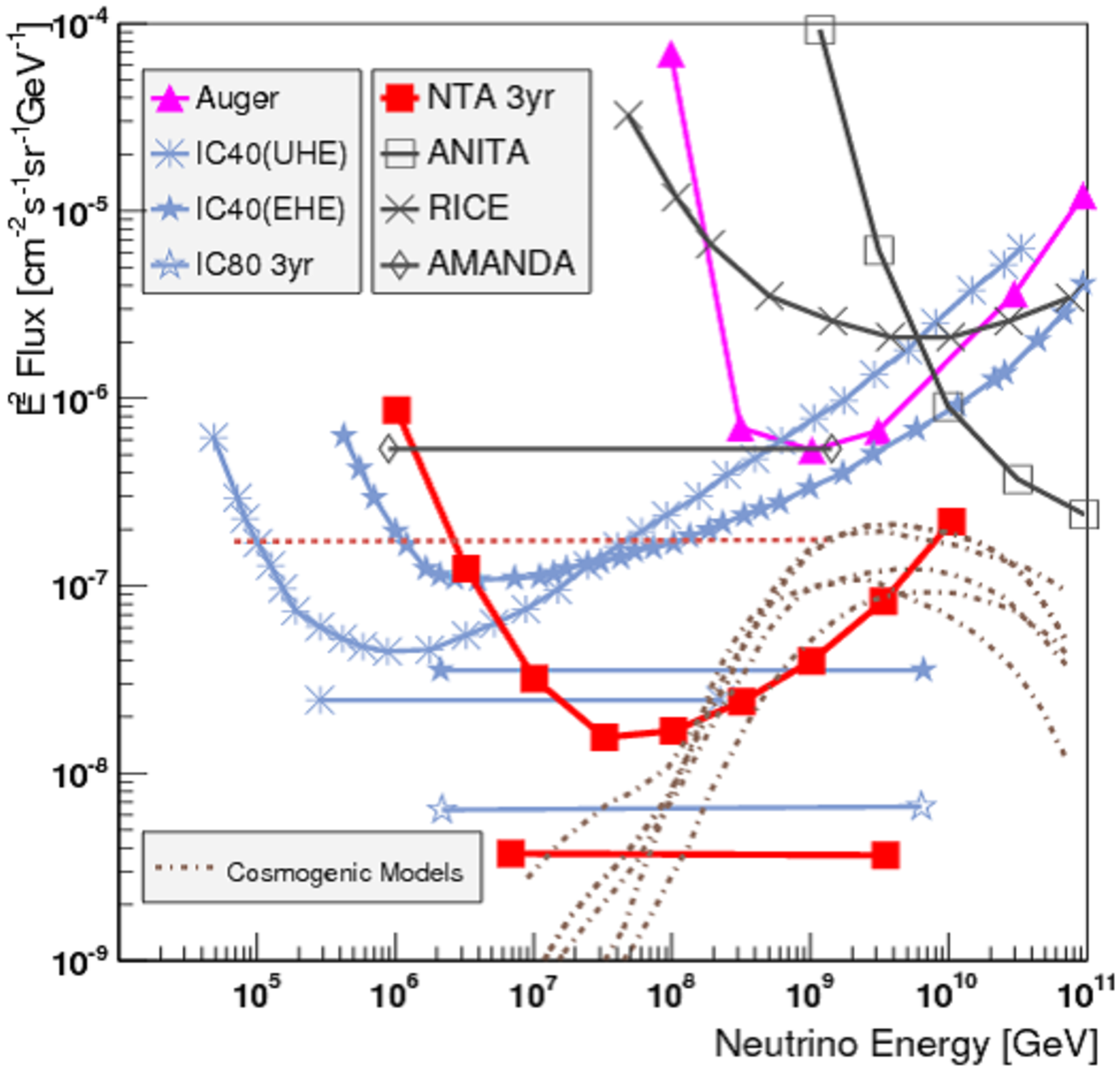}
\end{minipage}
\end{tabular}
\end{center}
 \caption{
(left) Differential sensitivities calculated as Feldman-Cousins 90\% CL limit
 for various LC designs and $\theta_{\rm elev} = -10^\circ$,
 in comparison with published sensitivities of IceCube and Auger,
 together with theoretical models (dashed lines for Ref.~\cite{Huemmer});
(right) Diffuse sensitivity of NTA with 3 year observation (maroon squares),
 together with cosmogenic neutrino flux models and related efforts.
}
 \label{fig:sensetivity}
\end{figure}

We show in Fig.~\ref{fig:sensetivity} (left) the
differential sensitivity for NTA, compared side by side with
IceCube and Auger capabilities (for illustration only).
The solid theoretical fluence curve is ruled out by IceCube,
but subsequent models for GRB neutrino flux, e.g. \cite{Huemmer},
can be probed by NTA.
Given the PeV neuntrino events observed by IceCube,
this search is now mandatory.
The power of NTA is to survey $\nu_{\tau}$ point source objects with
the best-yet sensitivity in the detection solid angle for $\nu_{\tau}$
defined as $-30^{\circ} < \theta_{\rm elev} < 0^{\circ}$ and
$0^{\circ} < \phi_{\rm azi} < 360^{\circ}$, and for
$10\ {\rm PeV} < E_{\nu_{\tau}} < 1\ {\rm EeV}$.
Fig.~\ref{fig:sensetivity} (left) shows that the NTA survey depth
can reach $z \lesssim 0.15$, of order 2 billion lightyears.
The location of NTA on Hawaii Island allows us to survey
the Galactic Center (GC) for more than several hundred hours each year.

NTA sensitivity for diffuse $\nu_{\tau}$ flux (maroon squares)
for 3 year ($\sim$ 9.5$\times$10$^{6}$~s) observation is
given in Fig.~\ref{fig:sensetivity} (right), assuming
duty cycle of 10\% and trigger conditions as above.
For lack of space, further discussion on angular resolution
and background simulation can be found in the LOI~\cite{LOI}. 

\section{Organization: Collaborators Welcome !!}

The IceCube Collaboration operating at the South Pole
consists of $\sim 250$ people, and 39 instiutions from 11 countries.
The corresponding numbers for the Auger Collaboration,
operating on the Pampas in Argentina,
are $\sim 480$ people from 18 countries.
We have estimated that a minimum of 30 Detector Units (DUs)
are needed for the coverage given in Table~\ref{tab:site} for NTA,
distributed over four mountain sites on Hawaii Island,
assuming the DU FOV to be $32^{\circ} \times 32^{\circ}$.
Each DU would require four Light Collectors (LCs),
one trigger and readout unit.
The estimated cost per DU, based on Ashra-1 experience, is $\sim$100M yen.
Allowing for some infrastructure and siting,
but not running and maintenance costs,
a crude cost estimate is 5000M~yen for the construction of NTA.
We estimate therefore, given the size and challenge of the project,
that NTA eventually would be a Collaboration consisting of
up to 10 countries.

As we design the NTA instruments and explore site options,
collaboration organization has started,
although one is still in the chicken\&egg phase:
one probably would need a major funding contribution
at more than 50\% total cost to start attracting
international collaborators,
and only then would one be able to seriously devise
a schedule towards the scientific goal.
Currently, we have a small International Executive Board (IEB)
of national representatives from Japan, Taiwan and U.S.
(with national affairs handled domestically),
and initial meetings started since late 2012,
but other groups are invited to join.
The time frame for the proposed project will be determined
both by budgetary and scientific considerations.
A workshop is planned for the first quarter of 2015 to discuss
the design and plans of the project with interested colleagues.
Major decisions on hardware implementation is expected in 2015
towards the Project Proposal.
Funding efforts would likely take 2 more years.
If Japanese core funding is received in time,
we hope to start experimental operations using
at least part of Site0 and Site1 by 2018;
Ashra-1 would continue to run for both testing and scientific purposes.
The expected construction time for full NTA would be of order 5 years.

In conclusion, the scientific goal of
\begin{center}
``\emph{Clear Discovery and Identification of Nonthermal Hadronic Processes \\
in the Universe, be it Galactic, Extragalatic, or Cosmogenic}.''
\end{center}
is reachable, and in view of the IceCube PeV astro-$\nu$ events,
it has become mandatory.
A Collaboration, the NTA, is needed to achieve this,
and collaborators are welcome!

\section{Epliogue: GZK or Astro?}

The discussion at VHEPA2014 has shed some light on
future prospects for VHE cosmic neutrinos.
The ``cosmogenic'' neutrino flux models given in
Fig.~\ref{fig:sensetivity} (right) are the famed GZK neutrinos,
which arise through the ``$\Delta$" conversion
by interaction of UHE ($10^{19}$ eV or above) cosmic protons
with the cosmic background radiation (in microwave).
Once seemingly assured, these flux estimates are in a state of flux.

At VHEPA2014, Jordan Hanson~\cite{Jordan} talked about various
radio detection efforts that are currently at the prototyping phase,
aiming ultimately at the GZK neutrino flux above $10^{18}$ eV.
Take ARIANNA for example, the full detector
3-year result can detect the nominal GZK flux.
However, because of the ``CR Composition'' issue,
this flux could drop by up to an order in magnitude!
It may be questioned whether ARIANNA, or any similar
detector, could have such reach.

This ``CR Composition'' issue arose through the
success of the Auger program, as presented by
Karl-Heinz Kampert~\cite{Karl-Heinz}, the current spokesperson,
at VHEPA2014. Auger data suggest that
the CR composition moves away from protons
towards iron, starting around $10^{18.6}$ eV,
resulting in the depletion of expected GZK flux.
Thus, although this is mildly disputed by the
TA Collaboration~\cite{Sagawa},
``GZK effect or exhausted sources?'' has become
\#1 driving question for the Auger Upgrade
which is currently under Scientific Review.

Auger has also studied its capabilities with UHE-$\nu$s,
as presented by Jaime Alvarez-Mu\~nez~\cite{Jaime} at VHEPA2014.
Because of the already fixed Auger detector configuration,
Auger's best sensitivity is around $10^{18}$ eV,
complementary to that of IceCube around $10^{15}$ eV.
But, the fact that IceCube sees Astro-$\nu$s at PeV
energy, one expects a spectrum to be probed between
IceCube and Auger sensitivities.
This is precisely where NTA fits in, with great
pointing accuracy to pick out point sources within
2 Glyrs. After all,
the target is not the diffuse source, such as GZK-$\nu$s,
but “nearby” astrophysical Point Sources.
Direct observation of such sources with the ES-$\nu_\tau$
technique would reveal the existence of hadronic acceleration
mechanisms, and open a new chapter in CR and astrophysics.


\end{document}